\documentclass[aps,reprint,floatfix]{revtex4-1}
\usepackage{latexsym}
\usepackage{amsfonts}
\usepackage{mathrsfs}
\usepackage{amssymb,amscd}
\usepackage[all]{xy}
\usepackage{amsmath}
\usepackage{amsthm}
\usepackage{fancyhdr}
\usepackage{fancybox}
\usepackage{xcolor}
\usepackage{hyperref}
\usepackage{verbatim}
\usepackage{makeidx}
\usepackage{indentfirst}
\usepackage[english]{babel}
\usepackage{graphicx}
\usepackage{listings}
\usepackage{subfigure}
\usepackage{multirow}
\usepackage{physics}
\usepackage{amsmath}
\usepackage{framed}
\usepackage{bbold}
\usepackage{siunitx}
\usepackage{cases}

\begin{document}

\title{Charge-to-spin conversion efficiency in ferromagnetic nanowires by spin torque ferromagnetic resonance: Reconciling lineshape and linewidth analysis methods}

\author{Jun-Wen Xu}
\author{Andrew D. Kent}
\email[]{andy.kent@nyu.edu}
\affiliation{Center for Quantum Phenomena, Department of Physics, New York University, New York 10003, USA}
\date{\today}

\begin{abstract}
Spin orbit torques are of great interest for switching the magnetization direction in nanostructures, moving skyrmions and exciting spin waves.
The standard method of determining their efficiency is by spin torque ferromagnetic resonance (ST-FMR), a technique that involves analyzing the resonance linewidth or lineshape.
On microstuctures these two analysis methods are quite consistent.
Here we present ST-FMR results on permalloy (Ni$_{80}$Fe$_{20}$) nanowires --- with widths varying from $150$ to $\SI{800}{nm}$ --- that show that the standard model used to analyze the resonance linewidth and lineshape give different results;
the efficiency appears greatly enhanced in nanowires when the lineshape method is used.
A ST-FMR model that properly accounts for the sample shape is presented and shows much better consistency between the two methods.
Micromagnetic simulations are used to verify the model.
These results and the more accurate nanowire model presented are of importance for characterizing and optimizing charge-to-spin conversion efficiencies in nanostructures.
\end{abstract}
\maketitle

\section{Introduction}
Spin orbit torques (SOTs) are being actively considered for using in the next generation memory devices for magnetization switching~\cite{Liu2012,Miron2011}, spin oscillators~\cite{Demidov2012,Duan2014} and racetrack memories, including those using magnetic skyrmions~\cite{Fert2013,Woo2017, montoya2018,jiang2016}.
SOTs are fundamentally based on charge-to-spin conversion by the spin-Hall effect~\cite{Hoffmann2013} or Rashba effect~\cite{Manchon2008}. 
These effects result in a spin current or spin accumulation that is transverse to the direction of charge current flow. 
An advantage of SOTs is that a charge current does not need to flow through the magnetic layer to switch its magnetization direction.
The torques can thus be used to switch magnetic insulators~\cite{Peng2016} or the free layer of a magnetic tunnel junction~\cite{Liu2012} without current flowing through the insulating layer. 

SOTs are principally of interest in nanostructured samples, samples with minimum dimension less than a micron.
In such samples the switching current density can be relatively large while the current is still small and the torques on the magnetization associated with charge current induced Oersted fields can be much smaller than the SOT.
However, the magnitude and form of the torques are most often characterized in micron scale samples.
The most widely used SOT characterization method is spin torque ferromagnetic resonance (ST-FMR)~\cite{Sankey2006,Tulapurkar2005}.
In this technique a radio frequency (rf) charge current leads to a rf spin torque that excites magnetization dynamics, with the largest response occurring at the ferromagnetic resonance (FMR) frequency.
The magnitude of the torque can be determined by either analyzing the linewidth of the response or the lineshape of the resonance~\cite{Petit,Liu2011}.
This allows extracting the effective charge-to-spin conversion efficiency, which is proportional to the spin-Hall angle $\theta_\mathrm{SH} = j_\mathrm{s}/j_\mathrm{c}$~\footnote{ST-FMR experiments relate the SOT to the charge current. The SOT depends on the spin current as well as other factors such as the interface transparency and the spin diffusion lengths in the materials.}.
The linewidth and lineshape  analysis methods have been shown to give consistent results when applied to microstructures~\cite{Liu2011,Zhang2015,Ralph2016,Ganguly2014}. 

In this paper we show that when ST-FMR results on nanostructures are analyzed with the standard ST-FMR model the results are generally not consistent;
with the lineshape analysis method the efficiency  appears greatly enhanced in nanowires.
We present ST-FMR results on permalloy nanowires with linewidths varying from 150 to $\SI{800}{nm}$ and a model that properly accounts for the sample geometry.
Our nanowire model provides much better consistency between the lineshape and linewidth analysis methods.

\section{Experiment}
\begin{figure}[ht!]
    \includegraphics{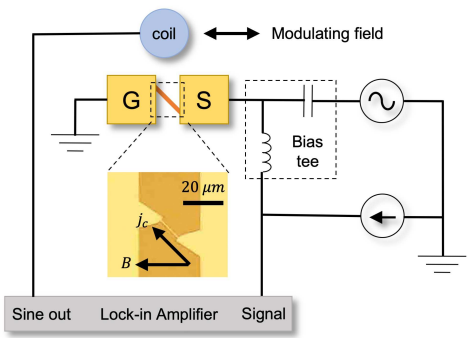}
    \caption{ST-FMR setup. The nanowire is oriented 45 degrees to the external field $B$. An rf and dc current are applied using a bias tee, while a small field coil is used to modulate the applied field at a low frequency.}
    \label{fig:setup}
\end{figure}

Experiments were conducted on Permalloy (Py, Ni$_{80}$Fe$_{20}$) nanowires fabricated from thin films.
The films are grown by dc magnetron sputtering on thermally oxidized silicon substrates with layer stack SiO$_2$/Ta (3)/Py(5)/Pt(6), where the numbers are the layer thicknesses in nm.
Ta is a seed layer~\cite{Nguyen2015} and the Pt layer is the main source of the spin current.
It also protects the Py from oxidation.
Electron beam lithography followed by argon ion milling is used to define the sample geometry.
We deposit Ti(5)/Au(50) by evaporation and liftoff for the contact pads.
An image of the nanowire and contact pads is shown in the inset of Fig.~\ref{fig:setup}. 
The width of nanowires varies from 150 to $\SI{800}{nm}$ with a fixed aspect ratio of 20. 

Figure~\ref{fig:setup} shows the ST-FMR setup. The nanowire is oriented at 45 degrees to the applied in-plane field in order to have a large ST-FMR signal.
A rf charge current is applied to the sample using a ground-signal (GS) probe connected to contact pads to the nanowire.
The rf current is input to the high frequency port of a bias tee, while the bias tee's low frequency port is connected to a dc current source and a lock-in amplifier. 

The measurement principal is as follows.
The rf current produces an rf torque on the ferromagnetic layer, causing this layer's magnetization to precess.
The largest precession amplitude occurs at the FMR frequency.
The oscillation in the magnetization leads to oscillations in the nanowire resistance due to the anisotropic magnetoresistance (AMR) of Py.
A dc voltage $V_\mathrm{mix}$ appears across the nanowire due to the mixing of the rf charge current and resistance oscillations at the same frequency.

In order to increase the signal-to-noise ratio, a small amplitude magnetic field of $\SI{0.2}{mT}$ (an amplitude much less than the FMR linewidth) is modulated at low frequency $\SI{727}{Hz}$, a frequency far less than the FMR frequency.
The lock-in amplifier is set to detect the signal at this low frequency.
The modulated voltage signal measured is $V(B) \propto \dd{V_\mathrm{mix}}/\dd{B}$ ~\cite{Goncalves2013}.
A single measurement corresponds to sweeping the external field from high to low values at fixed rf current amplitude and frequency.
The sweep is from high to low field to ensure each measurements starts from a saturated magnetic state.
For each rf frequency, the field is swept multiple times and the signal is averaged in order to further increase the signal-to-noise ratio.
The rf frequency is varied from 9 to $\SI{15}{GHz}$ and magnetic fields up to $\SI{0.3}{T}$ are applied.
The power of our rf source is fixed at $\SI{10}{dBm}$ and we have verified that the resonance response is in the linear regime.

The lineshape analysis method consists of analyzing the peak shape.
The peak in the lock-in voltage as a function of magnetic field, $V(B)$, can be decomposed into the sum of the derivative of Lorentzian and anti-Lorentzian functions (see the supplementary materials for further details):
\begin{equation}
    V(B)=\frac{-S(B-B_0)\Delta + A\left[(B-B_0)^2-(\Delta/2)^2\right]}{\left[(B-B_0)^2+(\Delta/2)^2\right]^2}
    \label{Eq:Spectra},
\end{equation}
where $B_0$ is the resonance field, $B$ is the external field in vacuum (i.e. $B=\mu_0 H$, where $\mu_0$ is the permeability of free space) and $\Delta$ is the resonance full width at half maximum (FWHM).
$S$ and $A$ are Lorentzian and anti-Lorentzian amplitudes, respectively.
In contrast, for the linewidth analysis method, a dc current is applied to the sample through the low frequency port of the bias tee.
The variation of the resonance peak linewidth with dc current is used to determine the efficiency.

In the following we denote the standard ST-FMR model the thin film model, as this model assumes the ferromagnet has an easy-plane magnetic anisotropy, with no preferred magnetization axis in the plane. 
The nanowire model we present considers the in-plane magnetic shape anisotropy that we discuss further below.

\subsection{Results and analysis with thin film model}
\begin{figure}[t!]
    \includegraphics{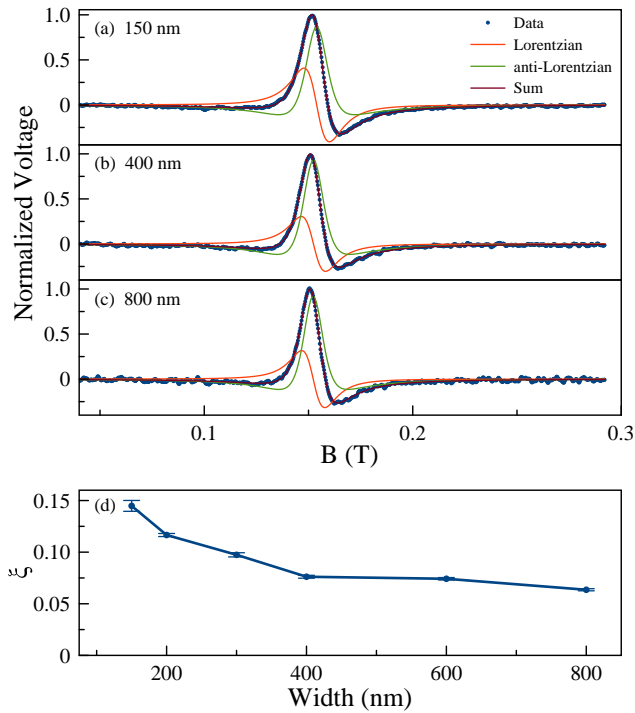}
    \caption{(a)-(c) Normalized spectra of 150, 400 and 800 nm width nanowires at 12 GHz. The normalization voltages are $\SI{2.43}{\mu V}$, $\SI{1.71}{\mu V}$ and $\SI{1.48}{\mu V}$ respectively. The red and green curves show the anti-Lorentzian and Lorentzian components of the resonance peak respectively. (d) The efficiency versus nanowire width determined from the lineshape using the thin film model.}
    \label{Fig:Res}
\end{figure}

Figure~\ref{Fig:Res}(a)-(c) show spectra of 150, 400 and $\SI{800}{nm}$ linewidth nanowires at a fixed rf frequency of $\SI{12}{GHz}$, where each spectra is normalized to its maximum value.
It is clear that the signal-to-noise ratio is more than adequate for detailed analysis.
We fit the spectra to Eq.~\ref{Eq:Spectra} to determine the Lorentzian $S$ and anti-Lorentzian $A$ amplitudes, the resonance field $B_0$ and the linewidth $\Delta$.

In the thin film model the ratio of the Lorentzian and anti-Lorentzian amplitudes is used to compute the charge-to-spin conversion efficiency, $\xi$, given by~\cite{Liu2011}:
\begin{equation}
    \xi = \frac{S}{A} \sqrt{1 + \frac{\mu_0 M_\mathrm{eff}}{B_0}} \frac{e \mu_0 M_\mathrm{s} t d}{\hbar}, 
    \label{Eq:ThinFilmModel}
\end{equation}
where $t$ is the thickness of ferromagnetic layer and $d$ is the thickness of heavy metal layer.
$M_\mathrm{eff}$ is the effective magnetization that characterizes the ferromagnet's easy plane anisotropy and $M_\mathrm{s}$ is its saturation magnetization, $e$ is the electron charge, and $\hbar$ is the reduced Planck's constant. 
We note that this model assumes a negligible field like SOT; it assumes that the field-induced torque is associated with the Oersted field from the charge current.
This is a reasonable assumption for our samples, as it has been reported that the field-like SOT decays with magnetic layer thickness, with a characteristic decay length of $\simeq \SI{2}{nm}$~\cite{Pai2015}.
Thus for our $\SI{5}{nm}$ thick ferromagnetic layer this torque will be far less than the anti-damping torque and the Oersted field induced torque.

Using the fitting parameters derived from the data in Fig.~\ref{Fig:Res} and taking $M_\mathrm{eff}=M_\mathrm{s}$, Eq.~\ref{Eq:ThinFilmModel} is used to determine $\xi$.
The result is shown in Fig.~\ref{Fig:Res}(d).
The efficiency increases dramatically (i.e. by more than a factor of 2) as the wire linewidth is reduced.
For larger nanowire widths ($\gtrsim 400$ nm), $\xi=0.65$ which is consistent with that reported value for Pt/Py interfaces ~\cite{Liu2011,Zhang2015,Ralph2016}.
It is reasonable to associate the main component of SOT with this interface because Ta is much more resistive \cite{Zhang2003} than Pt and thinner in our samples, so most of the current passes through the Pt layer.

\begin{figure}[t!]
    \includegraphics{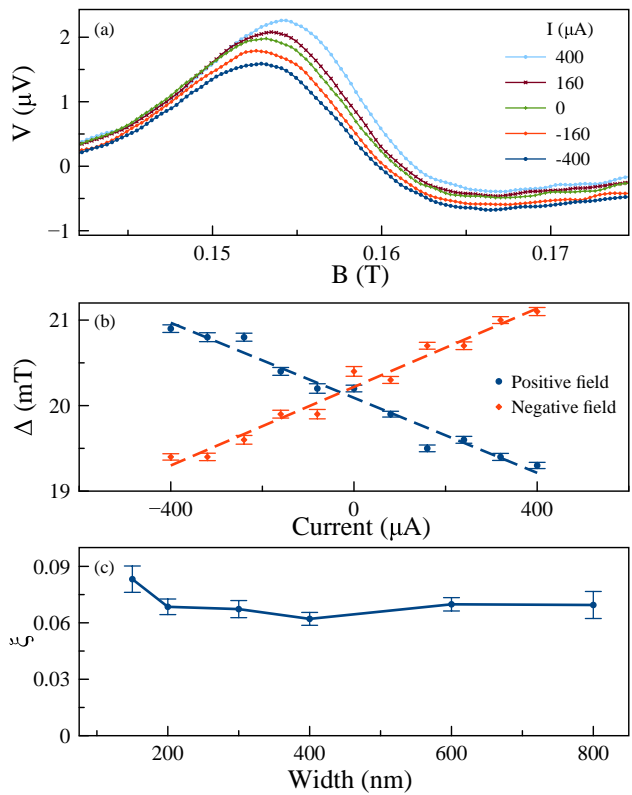}
    \caption{(a) ST-FMR spectra of a $\SI{300}{nm}$ width nanowire at $\SI{12}{GHz}$ at several dc bias currents. (b) Linewidth as a function of dc current for both positive and negative field polarities. (c) The efficiency as a function of the nanowire width determined using the linewidth analysis method.}
    \label{Fig:linewidth}
\end{figure}

We also studied the efficiency using the linewidth analysis method, expecting to see a similar trend with nanowire width.
As noted above, in this method a dc current is applied to the nanowire and the ST-FMR linewidth is determined as a function of the current.
For one polarity of the current the SOT opposes the damping and leads to a reduced ST-FMR linewidth, while for the opposite current, SOT increases the damping and the ST-FMR linewidth increases~\cite{STT}.
Figure~\ref{Fig:linewidth} shows selected spectra at fixed rf frequency for several dc bias currents. 
Fitting these spectra to Eq.~\ref{Eq:Spectra} we determine the linewidth as function of the dc bias.
The results are shown in Fig.~\ref{Fig:linewidth}(b).
For positive field, the linewidth decreases with increasing bias current, and vice versa.
The slope of the linewidth versus dc bias curves is $\SI{-2.2e-3}{mT/\mu A}$ for positive field and $\SI{2.3e-3}{mT/ \mu A}$ for negative field.

In the thin film model the efficiency is related to the slope of the ST-FMR resonance linewidth versus dc bias current~\cite{Liu2011, Petit}: 
\begin{equation}
    \xi = \frac{\gamma e \left(B_0 + \mu_0 M_\mathrm{eff}/2 \right) M_\mathrm{s} t}{2 \pi \hbar f \sin \phi} \dv{\Delta}{j_\mathrm{c}},
    \label{Eq:Thinfim_width}
\end{equation}
where $j_\mathrm{c}$ is the current density in the Pt layer and $\phi$ is the angle between charge current and external magnetic field.
Figure~\ref{Fig:linewidth}(c) shows the resulting $\xi$ versus nanowire width; $\xi$ varies between 0.062 - 0.077.
Comparing Fig.~\ref{Fig:Res}(d) and Fig.~\ref{Fig:linewidth}(c) it is clear there is a significant discrepancy in the SOT efficiencies deduced from these models.
It is also clear that some basic physics in the modeling of the efficiency in nanowires is not captured by the standard thin film model.

\subsection{Nanowire model}
\begin{figure}[b]
    \includegraphics{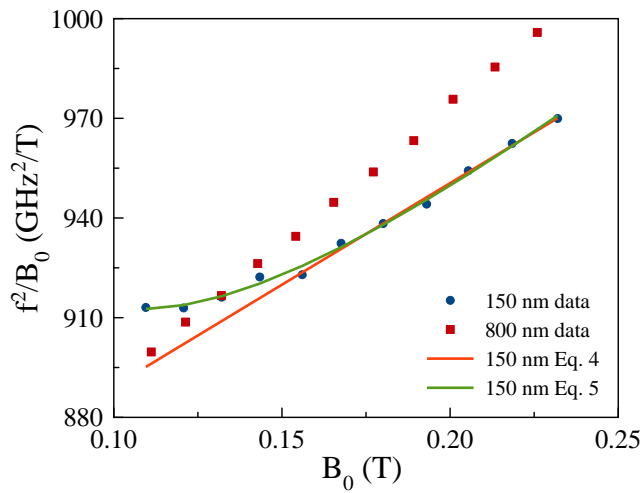}
    \caption{Resonance field versus rf frequency for 150 and $\SI{800}{nm}$ nanowires, fit with Eq.~\ref{Eq:f2H} and Eq.~\ref{Eq:Kittel_new}}
    \label{Fig:Kittel}
\end{figure}

We begin by analyzing the dependence of resonance field on rf frequency to test a basic assumption of the thin film model.
The model assumes an easy-plane magnetic anisotropy with a resonance frequency-field relation: 
\begin{equation}
    \frac{f^2}{B_0} = \left(\frac{\gamma}{2 \pi} \right)^2 \left(B_0 + \mu_0 M_\mathrm{eff} \right) \label{Eq:f2H}.
\end{equation}
That is $f^2/B_0$ should be a straight line when plotted versus field $B_0$, with a slope proportional to the gyromagnetic ratio squared and intercept proportional to $M_\mathrm{eff}$.
The experimental data are shown in Fig.~\ref{Fig:Kittel}.
It is clear that the $\SI{800}{nm}$ data follows the expectations of the model.
However, the data for the $\SI{150}{nm}$ width wire deviates strongly from the straight line trend (orange curve) for resonance fields less than about $\SI{0.15}{T}$.

Assuming that applied fields greater than $\SI{0.15}{T}$ are required to fully saturate the magnetization of the $\SI{150}{nm}$ width wire, one may omit the lower resonance field data from the fits to Eq.~\ref{Eq:f2H} (see the orange line in Fig.~\ref{Fig:Kittel}).
Following this approach for all samples we find the fitting parameters $M_\mathrm{eff}$ and $\gamma/(2 \pi)$  given in Table~\ref{Table:ThinFilm}. 
For wide nanowires the parameters are those expected for Py, $\mu_0 M_\mathrm{eff} \approx \mu_0 M_\mathrm{s}=\SI{1}{T}$ and $\gamma/(2 \pi)=\SI{28}{GHz/T}$.
However, for the narrower samples, $\mu_0 M_\mathrm{eff}$ is larger and $\gamma/(2 \pi)$ is smaller.
This does not make physical sense as $\mu_0 M_\mathrm{eff}$ is associated with the demagnetization field perpendicular to the film plane, which decreases as the wire width decreases.
Without an additional easy plane magnetic anisotropy (e.g. associated with spin-orbit interactions at the Py interfaces), $M_\mathrm{eff}\leq M_\mathrm{s}$.  For example, for the $\SI{150}{nm}$ width nanowire, the demagnetization field in the width direction $\simeq \SI{0.05}{T}$~\cite{demag}, which is not negligible compared to the applied magnetic field in our experiments (0.1 to $\SI{0.3}{T}$).

\begin{table}[t]
    \centering
    \begin{tabular}{l|c c c c c c}
        \hline
        \hline
        $\text{Width}$ ($\SI{}{nm}$)        & 150    & 200   & 300   & 400   & 600   &800\\ 
        \hline
        $\mu_0 M_\mathrm{eff}$ ($\SI{}{T}$)   & \SI{1.35 \pm 0.05}{}   & \SI{1.19 \pm 0.01}{}   & \SI{1.19 \pm 0.02}{}   & \SI{1.11 \pm 0.02}{}   & \SI{1.04 \pm 0.01}{}   & \SI{0.96 \pm 0.01}{}\\ 
        $\gamma / 2 \pi$ ($\SI{}{GHz/T}$)   & \SI{24.7 \pm 0.4}{}   & \SI{26.2 \pm 0.1}{}   & \SI{26.3 \pm 0.2}{}   & \SI{27.2 \pm 0.2}{}   & \SI{28.0 \pm 0.1}{}   & \SI{28.9 \pm 0.1}{}\\
        \hline
        \hline
    \end{tabular}
    \caption{Effective magnetization and gyromagnetic ratio from fits to Eq.~\ref{Eq:f2H} for different width nanowires.}
    \label{Table:ThinFilm}
\end{table}

It is thus clear that the in-plane magnetic anisotropy associated with the sample shape is not negligible and needs to be considered in the analysis of the resonance field. The shape anisotropy can be described by a in-plane uniaxial anisotropy field $B_A$ parallel to the wire axis. The resulting resonance field condition becomes:
\begin{equation}
    f = \frac{\gamma}{2 \pi} \sqrt{\left(B_0 +B_A \right) \left(B_0 + B_A + \mu_0 M_\mathrm{eff} \right)} 
\label{Eq:Kittel_new}.
\end{equation}
The additional field $B_A$ is associated with the average demagnetization field in the width direction, i.e. the in-plane magnetic shape anisotropy. 
The fit to Eq.~\ref{Eq:Kittel_new} is shown as the green curve in Fig. \ref{Fig:Kittel}.
The fit accurately captures both the low and high field resonance data with the fit parameters for all the nanowires shown in Table~\ref{Table:M_new}. 
We now see, as expected, that $M_\mathrm{eff}$ decreases as the wire width decreases. Further, the gyromagnetic ratio is nearly independent of wire width.
\begin{table}[t]
    \centering
    \begin{tabular}{l|c c c c c c}
        \hline
        \hline
        $\text{Width}$ ($\SI{}{nm}$)        & 150    & 200   & 300  & 400   & 600   &800\\ 
        \hline
        $\mu_0 M_\mathrm{eff}$ ($\SI{}{T}$)   & \SI{0.80 \pm 0.08}{}  & \SI{0.83 \pm 0.07}{}   & \SI{0.94 \pm 0.02}{}   & \SI{0.94 \pm 0.02}{}   & \SI{0.96 \pm 0.04}{}   & \SI{0.92 \pm 0.05}{}\\
        $B_A$ ($\SI{}{mT}$)                  & \SI{14 \pm 3}{}& \SI{10 \pm 2}{} &\SI{7 \pm 1}{}    &\SI{3 \pm 1}{} &\SI{1 \pm 1}{} &\SI{1 \pm 1}{}\\
        $\gamma / 2 \pi$ ($\SI{}{GHz/T}$)   & \SI{29.6 \pm 1.0}{}   & \SI{29.6 \pm 0.8}{}   & \SI{28.7 \pm 0.2}{}   & \SI{28.8 \pm 0.2}{}   & \SI{28.9 \pm 0.5}{}   & \SI{29.4 \pm 0.6}{}\\
        \hline
        \hline
    \end{tabular}
    \caption{Effective magnetization, in-plane anisotropy field and gyromagnetic ratio from fits to Eq.~\ref{Eq:Kittel_new} for different width nanowires.}
    \label{Table:M_new}
\end{table}

As a consequence of the shape anisotropy the demagnetization field is not collinear with the applied field in our experiment.
Thus the precession axis of the magnetization in resonance is also no longer the applied field direction.
As a result the angle between the precession axis and the current is less than 45 degrees, $\phi < \pi/4$.
$\phi$ can be calculated numerically by solving
\begin{equation}
B \sin(\pi/4) =  N_x \mu_0 M_\mathrm{s} \sin \phi + B \tan \phi \cos(\pi/4),
\label{Eq:angle}
\end{equation}
where $N_x$ is the demagnetization coefficient in the width direction (see supplementary part 2).

Further, the charge current induced Oersted field for nanowires is different from the film case as well.
The in-plane Oersted field decreases at the edge of the wire, as its direction becomes more out of the film plane.
The ST-FMR model assumes a uniform magnetization response, so the average in-plane Oersted field is considered.
The average in-plane Oersted field for a very wide strip ($B_\mathrm{Oe} =\mu_0 J d/2$) is reduced, multiplied by a factor $\bar\Theta(\epsilon)<1$
(see supplementary part 3):
\begin{equation}
    \bar\Theta(\epsilon) = \frac{1}{\pi} \left[ 2 \arctan \epsilon - \frac{1}{\epsilon} \ln(1 + \epsilon^2)\right],
\end{equation}
where $\epsilon \equiv w/d$ is the ratio of width to the thickness of the Pt layer. In the film case, $\epsilon \rightarrow \infty$, $\bar{\Theta}=1$.
For a $\SI{150}{nm}$ nanowire, $\epsilon = 25$ and $\bar{\Theta} \approx 0.89$.

In the nanowire model lineshape analysis method:
\begin{equation}
    \xi = \frac{S}{A} \sqrt{1 + \frac{\mu_0 M_\mathrm{eff}}{B_0 + B_A}} \frac{e \mu_0 M_\mathrm{s} t d}{\hbar} \bar\Theta.
    \label{Eq:shape_new} 
\end{equation}
While in the linewidth analysis method, the expression is the same as that in Eq.~\ref{Eq:Thinfim_width}. However, $\phi$ now needs to be evaluated numerically using Eq.~\ref{Eq:angle}.

\begin{figure}[t]
    \includegraphics{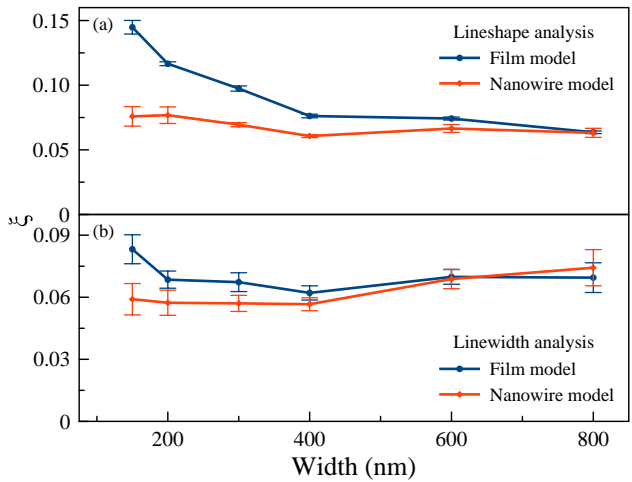}
    \caption{Comparison of the thin film and nanowire models using (a) the lineshape analysis method and (b) the linewidth analysis method. \label{fig.final}}
\end{figure}

Figure~\ref{fig.final} shows the ST-FMR data analysis using our nanowire model. The efficiency $\xi$ using both lineshape analysis and linewidth analysis methods for all width of the nanowires is between 0.6-0.7. Also, there is no longer a significant enhancement in $\xi$ at small nanowire widths with the lineshape analysis method.

\subsection{Nanowire lineshape versus linewidth analysis methods}

We find that the nanowire lineshape and linewidth analysis methods are in reasonable accord with small differences, about $10$ \% differences in $\xi$, comparable to the measurement error.
There are reasons that the lineshape method may be more accurate.
First, this is because the lineshape analysis method is self-calibrated; it is a measure of the ratio of the SOT to the Oersted field torque which are both proportional to the current density in the heavy metal layer.
Hence, the ratio of Lorentzian and anti-Lorentzian peak amplitude is independent of the charge current density.
In contrast, in the linewidth analysis method, it is important to accurately estimate the current density passing through the heavy metal layer.
Second, in the linewidth analysis method, the angle between the magnetization precession axis and charge current is estimated assuming a uniform demagnetization field for a cuboid shaped sample~\cite{demag}, which is an approximation (see the last part of Sec.~\ref{Sec:MM}).
Third, during a field swept spectrum, this angle changes with the external field.
Here we take the angle, at the resonance field, as an approximation, which leads to an additional error.
Finally, the most important set of data for the linewidth analysis method is the variation of the linewidth with the dc bias current.
In our study, this change is only $\SI{1}{mT}$ when the current density varies by $\SI{\pm 2e11}{A/m^2}$, which is comparable to the $\SI{0.2}{mT}$ modulation field.
This introduces a systematic error in determining the linewidth variation with dc current.
Furthermore, the external field control is of order $\SI{0.01}{mT}$, limiting the signal-to-noise ratio.

\section{Micromagnetic simulations}
\label{Sec:MM}
In order to verify our nanowire model, we carried out micromagnetic modeling using MuMax (see supplementary part 4 for our code)~\cite{MUMAX,MUMAX2}.
We simulate nanowires with a width of 160, 320, 640 and $\SI{1280}{nm}$ and an aspect ratio of 16.
As in the experiment, the magnetization is driven by an oscillating Oersted field torque and SOT associated with a charge current.
After $\SI{4}{ns}$ of simulation time, the spatial averaged magnetization amplitude takes the form:
\begin{equation}
    m_x(t) = m_x \cos(2 \pi f t - \psi),
\end{equation}
where $m_x(t)$ is the instantaneous $x$ component of magnetization, $m_x$ is the amplitude of oscillation and $\psi$ is the phase delay between the magnetization response and rf current.
We sweep the external field $B$ at a fixed rf current amplitude and frequency.
The ST-FMR signal results from mixing of the rf current and anisotropic magnetoresistance:
\begin{equation}
    \begin{split}
        V_\mathrm{mix} =& \frac{1}{T} \int I(t)R(t) \dd{t}\\
        =& - \left(I \delta R \sin \phi \cos \phi \right) m_x \cos \psi, 
    \end{split}
\end{equation}
where the integral is over one period of the rf signal $T=1/f$ and $m_x \cos \psi$ is a sum of a Lorentzian and anti-Lorentzian function (see supplementary part 1).
Therefore, we fit the simulation result $m_x \cos\psi$ versus $B$ using the expression
\begin{equation}
    m_x \cos\psi = \frac{S(\Delta/2) + A \left( B - B_0 \right)}{\left( B - B_0 \right)^2 + \left(\Delta/2 \right)^2}
    \label{eq.Lorentzian},
\end{equation}
in order to extract the resonance field ($B_0$) the linewidth ($\Delta$), the Lorentzian amplitude ($S$) and the anti-Lorentzian amplitude ($A$), as we do for the experimental data.

Fitting the resulting resonance frequency versus magnetic field with Eq.~\ref{Eq:f2H} (the thin film model) the trends of $M_\mathrm{eff}$ and $\gamma/(2\pi)$ versus wire width are similar to those found for the experimental data (c.f. Table~\ref{Table:ThinFilm} and supplementary part 5 Table 1  for the simulation results); $M_\mathrm{eff}$ increases with decreasing wire width.
Since in the simulations $\mu_0 M_\mathrm{s} = \SI{1}{T}$ and $\gamma/(2 \pi) = \SI{28.0}{GHz/T}$, this discrepancy is unphysical.
However, using the nanowire expression, Eq.~\ref{Eq:Kittel_new}, gives reasonable values of $M_\mathrm{eff}$ and $\gamma/(2\pi)$ (see supplementary part 5 Table 2 for the simulation results).

\begin{figure}[ht!]
    \includegraphics{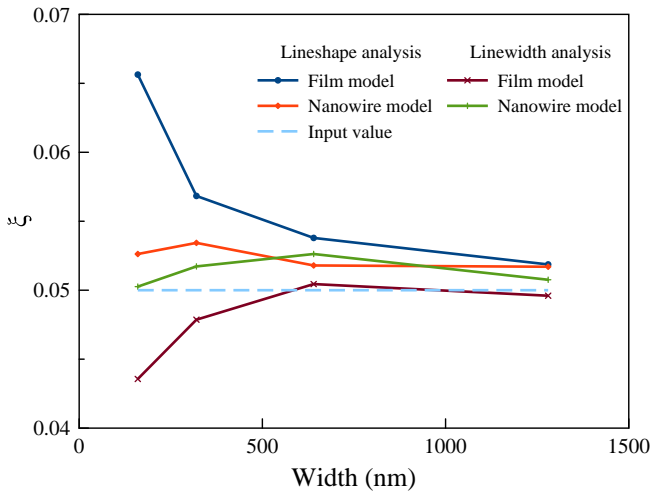}
    \caption{Simulation results of efficiency found using lineshape and linewidth analysis methods with both the thin film and nanowire models with for different wire widths. The dashed curve is the value of the efficiency used in the simulation, $\xi=0.05$.} \label{fig:simulation}
\end{figure}

The efficiency extracted from the models is plotted in Fig.~\ref{fig:simulation}. Note that this should be compared to the efficiency input into the micromagnetic simulations, $\xi=0.05$.
In the wide wire limit, both the thin film model and nanowire model are in good agreement using the lineshape analysis and linewidth analysis methods.
However, when the wire width becomes narrower, the thin film lineshape analysis gives a $\xi$ that increases (Fig.~\ref{fig:simulation}), as seen in experiment (Fig.~\ref{Fig:Res}(d)). Using the nanowire model $\xi$ is closer to 0.05.
The origin of the enhancement of $\xi$ is now clear. It is associated with the thin film model overestimating $M_\mathrm{eff}$ and the Oersted field, $\bar{\Theta}$.

In the linewidth analysis method, the thin film model shows that $\xi$ decreases in narrower wires.
But in the nanowire model, there is no longer such a decrease.
We have not seen such a decrease in our experimental results, Fig.~\ref{Fig:linewidth}(c). This is because in the thin film model (Eq.~\ref{Eq:Thinfim_width}), we overestimate $M_\mathrm{eff}$ while we underestimate $1/\sin \phi$, which compensate each another.

Finally, as discussed near Eq.~\ref{Eq:angle}, the equilibrium magnetization angle $\phi$ was calculated with demagnetization coefficients associated with a uniformly magnetized cuboid.
To estimate the error associated with this assumption, we compared the average magnetization angle determined in micromagnetics simulations from that found assuming a uniformly magnetized cuboid.
For the latter, the demagnetization field is  calculated following Ref.~\cite{demag} and input into a micromagnetics simulation as a fixed uniform field (turning off the internal demagnetization field in the simulation, see supplementary part 6 for details).
The relation of $B$ and $\sin \phi$ both using micromagnetics and a uniform demagnetization field with different width of nanowires is plotted in supplementary part 6.
We find that assuming a uniform demagnetization and following our procedure using Eq.~\ref{Eq:angle} produces a negligible error in the nanowire linewidth analysis method.

\section{Summary}
In summary we have introduced an analytic ST-FMR model for nanowires that accounts for their shape in a straightforward way.
The model gives reliable efficiency results either by analyzing the ST-FMR lineshape or the ST-FMR linewidth as a function of current.
As the primary interest in spin orbit torques is in exciting magnetization dynamics and switching the magnetization of nanostructures, our model can be of importance in reliable characterization and optimizing charge-to-spin conversion efficiencies in such structures.
Further, the model can be extended to other sample shapes and to include additional physics associated, for example, with the quantization of spin wave modes in nanostructures and spin-pumping.

\begin{acknowledgments}
We thank Dr. Christopher Safranski for discussions of ST-FMR analysis methods and Dr. Nahuel Statuto for the guidance in the micromagnetic modeling. 
This research was supported by DARPA Grant No. D18AP0000 and the National Science Foundation under Grant No. DMR-1610416. The nanostructures were realized at the Advanced Science Research Center NanoFabrication Facility of the Graduate Center at the City University of New York.
\end{acknowledgments}

\bibliographystyle{apsrev4-1}
\bibliography{main}

\end{document}